\documentclass{aastex61}

\shorttitle{unexpected cyclicity in cosmic ray protons}
\shortauthors{Adriani et al.}
\usepackage{amsmath}

\begin{document}

\title{Unexpected cyclic behavior in cosmic ray protons observed by PAMELA at 1 AU.}

\correspondingauthor{M. Ricci}
\email{marco.ricci@lnf.infn.it}

\author{O. Adriani} 
\affiliation{University of Florence, Department of Physics, I-50019 Sesto Fiorentino, Florence, Italy}
\affiliation{INFN, Sezione di Florence, I-50019 Sesto Fiorentino, Florence, Italy}

\author{G. C. Barbarino} 
\affiliation{University of Naples ``Federico II'', Department of Physics, I-80126 Naples, Italy}  
\affiliation{INFN,Sezione di Naples, I-80126 Naples, Italy}

\author{G. A. Bazilevskaya} 
\affiliation{Lebedev Physical Institute, RU-119991, Moscow, Russia}

\author{R. Bellotti}
\affiliation{University of Bari, Department of Physics, I-70126 Bari, Italy}
\affiliation{INFN, Sezione di Bari, I-70126 Bari, Italy}

\author{M. Boezio}
\affiliation{INFN, Sezione di Trieste, I-34149 Trieste,Italy} 

\author{E. A. Bogomolov}
\affiliation{Ioffe Physical Technical Institute, RU-194021 St. Petersburg, Russia} 

\author{M. Bongi} 
\affiliation{University of Florence, Department of Physics, I-50019 Sesto Fiorentino, Florence, Italy}
\affiliation{INFN, Sezione di Florence, I-50019 Sesto Fiorentino, Florence, Italy}

\author{V. Bonvicini} 
\affiliation{INFN, Sezione di Trieste, I-34149 Trieste,Italy}

\author{A. Bruno} 
\affiliation{University of Bari, Department of Physics, I-70126 Bari, Italy}

\author{F. Cafagna} 
\affiliation{INFN, Sezione di Bari, I-70126 Bari, Italy}

\author{D. Campana} 
\affiliation{INFN,Sezione di Naples, I-80126 Naples, Italy} 

\author{P. Carlson} 
\affiliation{KTH, Department of Physics, and the Oskar Klein Centre for Cosmoparticle Physics,AlbaNova University Centre, SE-10691 Stockholm, Sweden}

\author{M. Casolino}
\affiliation{INFN, Sezione di Rome ``Tor Vergata'', I-00133 Rome, Italy}  
\affiliation{RIKEN, EUSO team Global Research Cluster, Wako-shi, Saitama, Japan}

\author{G. Castellini} 
\affiliation{IFAC, I-50019 Sesto Fiorentino, Florence, Italy}

\author{C. De Santis}
\affiliation{INFN, Sezione di Rome ``Tor Vergata'', I-00133 Rome, Italy}  

\author{V. Di Felice} 
\affiliation{INFN, Sezione di Rome ``Tor Vergata'', I-00133 Rome, Italy}  
\affiliation{Space Science Data Center - Agenzia Spaziale Italiana, via del Politecnico, s.n.c., I-00133, Roma, Italy}

\author{A. M. Galper}
\affiliation{MEPhI: National Research Nuclear University MEPhI, RU-115409, Moscow, Russia} 

\author{A. V. Karelin}
\affiliation{MEPhI: National Research Nuclear University MEPhI, RU-115409, Moscow, Russia} 

\author{S. V. Koldashov} 
\affiliation{MEPhI: National Research Nuclear University MEPhI, RU-115409, Moscow, Russia} 

\author{S. Koldobskiy} 
\affiliation{MEPhI: National Research Nuclear University MEPhI, RU-115409, Moscow, Russia} 

\author{S. Y. Krutkov} 
\affiliation{Ioffe Physical Technical Institute, RU-194021 St. Petersburg, Russia}

\author{A. N. Kvashnin}
 \affiliation{Lebedev Physical Institute, RU-119991, Moscow, Russia}

\author{A. Leonov} 
\affiliation{MEPhI: National Research Nuclear University MEPhI, RU-115409, Moscow, Russia} 

\author{V. Malakhov} 
\affiliation{MEPhI: National Research Nuclear University MEPhI, RU-115409, Moscow, Russia} 

\author{L. Marcelli} 
\affiliation{INFN, Sezione di Rome ``Tor Vergata'', I-00133 Rome, Italy}  

\author{M. Martucci} 
\affiliation{University of Rome ``Tor Vergata'', Department of Physics, I-00133 Rome, Italy}
\affiliation{INFN, Laboratori Nazionali di Frascati, Via Enrico Fermi 40, I-00044 Frascati, Italy}

\author{A. G. Mayorov} 
\affiliation{MEPhI: National Research Nuclear University MEPhI, RU-115409, Moscow, Russia} 

\author{W. Menn}
\affiliation{Universitat Siegen, Department of Physics, D-57068 Siegen, Germany}

\author{M. Merg\`e} 
\affiliation{INFN, Sezione di Rome ``Tor Vergata'', I-00133 Rome, Italy}  
\affiliation{University of Rome ``Tor Vergata'', Department of Physics, I-00133 Rome, Italy}

\author{V. V. Mikhailov} 
\affiliation{MEPhI: National Research Nuclear University MEPhI, RU-115409, Moscow, Russia} 

\author{E. Mocchiutti} 
\affiliation{INFN, Sezione di Trieste, I-34149 Trieste,Italy} 

\author{A. Monaco} 
\affiliation{University of Bari, Department of Physics, I-70126 Bari, Italy}
\affiliation{INFN, Sezione di Bari, I-70126 Bari, Italy}

\author{N. Mori} 
\affiliation{INFN, Sezione di Florence, I-50019 Sesto Fiorentino, Florence, Italy} 

\author{R. Munini} 
\affiliation{INFN, Sezione di Trieste, I-34149 Trieste,Italy} 
\affiliation{University of Trieste, Department of Physics, I-34147 Trieste, Italy}

\author{G. Osteria}
\affiliation{INFN,Sezione di Naples, I-80126 Naples, Italy}

\author{B. Panico} 
\affiliation{INFN,Sezione di Naples, I-80126 Naples, Italy}

\author{P. Papini} 
\affiliation{INFN, Sezione di Florence, I-50019 Sesto Fiorentino, Florence, Italy} 

\author{M. Pearce}
\affiliation{KTH, Department of Physics, and the Oskar Klein Centre for Cosmoparticle Physics,AlbaNova University Centre, SE-10691 Stockholm, Sweden}

\author{P. Picozza} 
\affiliation{INFN, Sezione di Rome ``Tor Vergata'', I-00133 Rome, Italy}  
\affiliation{University of Rome ``Tor Vergata'', Department of Physics, I-00133 Rome, Italy}

\author{G. Pizzella} 
\affiliation{INFN, Laboratori Nazionali di Frascati, Via Enrico Fermi 40, I-00044 Frascati, Italy}

\author{M. Ricci}
\affiliation{INFN, Laboratori Nazionali di Frascati, Via Enrico Fermi 40, I-00044 Frascati, Italy}

\author{S. B. Ricciarini}
\affiliation{INFN, Sezione di Florence, I-50019 Sesto Fiorentino, Florence, Italy} 
\affiliation{IFAC, I-50019 Sesto Fiorentino, Florence, Italy}

\author{M. Simon} 
\affiliation{Universitat Siegen, Department of Physics, D-57068 Siegen, Germany}

\author{R. Sparvoli}
\affiliation{INFN, Sezione di Rome ``Tor Vergata'', I-00133 Rome, Italy} 
\affiliation{University of Rome ``Tor Vergata'', Department of Physics, I-00133 Rome, Italy}

\author{P. Spillantini}
\affiliation{MEPhI: National Research Nuclear University MEPhI, RU-115409, Moscow, Russia} 
\affiliation{Istituto Nazionale di Astrofisica, Fosso del cavaliere 100, 00133 Roma, Italy} 

\author{Y. I. Stozhkov} 
 \affiliation{Lebedev Physical Institute, RU-119991, Moscow, Russia} 

\author{A. Vacchi}
\affiliation{INFN, Sezione di Trieste, I-34149 Trieste,Italy} 
\affiliation{University of Udine, Department of Mathematics, Computer Science and Physics Via delle Scienze, 206, Udine, Italy}

\author{E. Vannuccini}
\affiliation{INFN, Sezione di Florence, I-50019 Sesto Fiorentino, Florence, Italy} 

\author{G. Vasilyev} 
\affiliation{Ioffe Physical Technical Institute, RU-194021 St. Petersburg, Russia} 

\author{S. A. Voronov} 
\affiliation{MEPhI: National Research Nuclear University MEPhI, RU-115409, Moscow, Russia} 

\author{Y. T. Yurkin} 
\affiliation{MEPhI: National Research Nuclear University MEPhI, RU-115409, Moscow, Russia} 

\author{G. Zampa}
\affiliation{INFN, Sezione di Trieste, I-34149 Trieste,Italy} 

\author{N. Zampa}
\affiliation{INFN, Sezione di Trieste, I-34149 Trieste,Italy} 

\begin{abstract}
Protons detected by the PAMELA experiment in the period 2006-2014 have been analyzed in the energy range between 0.40-50 GV to explore possible  periodicities besides the well known solar undecennial modulation. An unexpected clear and  regular feature has been found at rigidities below 15 GV, with a quasi-periodicity of $\sim$450 days. A possible Jovian origin of this periodicity has been investigated in different ways. The results seem to favor a small but not negligible contribution to cosmic rays from the Jovian magnetosphere, even if other explanations cannot be excluded. 
\end{abstract}

\keywords{astroparticle physics --- 
cosmic rays --- Sun: heliosphere}

\section{Introduction} \label{sec:intro}

The dominant and most important time scale in cosmic rays, related to solar activity, is the 11-year cycle \cite{Tobias2002}. This quasi-periodicity is translated into the galactic cosmic rays (GCRs) intensities widely recorded by the network of ground stations since the 1950s \cite{Lockwood1967}.
Later, a 22-year cycle was discovered, linked to the reversal of the Heliospheric Magnetic Field (HMF) taking place during each period of large solar activity \cite{Webber1988}.
There are also indications of GCRs periodicities of 50-65 years, 90-130 years and also for a periodicity of more than 200 years \cite{Potgieter2013}. 

Moreover, short periodicities (like 25-27 days and 1-day cycles) have been observed in many GCR data, related to the rotation of the Sun and of the Earth respectively \cite{Alania2011,Schwachheim1960}. 
How these periodicities are generated and how they could be explained through the not fully established Parker heliospheric coefficients \cite{Strong2007} is still a matter of study. More recently, new data have been provided by the high-precision measurements of GCRs performed by the satellite experiment PAMELA in a wide range of energy in the period 2006-2015, covering the end phase of the 23$^{rd}$ solar cycle and almost the whole 24$^{th}$ cycle (\cite{Adriani2013};\cite{Adriani2015};\cite{Adriani2016}).

\section{PAMELA detector}

The PAMELA apparatus consists of a combination of detectors capable of identifying particles up to Oxygen and giving information on charge, mass, rigidity and velocity from a few tens of MeV up to 1 TeV. The instrument is built around a permanent magnet with a silicon microstrip tracker, providing charge and track deflection information. A scintillator system provides trigger, time-of-flight and additional charge information. A silicon-tungsten calorimeter is used to perform general hadron-lepton separation. An anti-coincidence system of plastic scintillators allows the rejection of spurious events in the off-line phase.
A comprehensive description of the instrument, the mission profile, the scientific objectives and the results achieved during the PAMELA 10-year operation in space, can be found in \cite{Adriani2014}.

\section{The analysis}

The PAMELA proton data have been analyzed in three different rigidity ranges: 0.4-0.65 GV, 0.65-15 GV and 15-50 GV to explore possible unexpected periodicities. Following the methods described in \cite{Adriani2013} for a clear identification of low energy GCR protons, daily-averaged intensities have been calculated for the overall period July 9$^{th}$ 2006 - August 31$^{st}$ 2014. 

To ensure a clean galactic sample, contamination of solar particles from solar flares has been avoided discarding bunches of data taken during major Solar Particle Events (a list of these events is reported in \cite{Adriani2017}). Other short-term effects on GCRs, like Forbush decreases, caused by Coronal Mass Ejections passing through Earth, have been removed, even if their impact on the proton intensities appears negligible for most events. 

\begin{figure}
  \centering
    \includegraphics[width=\textwidth]{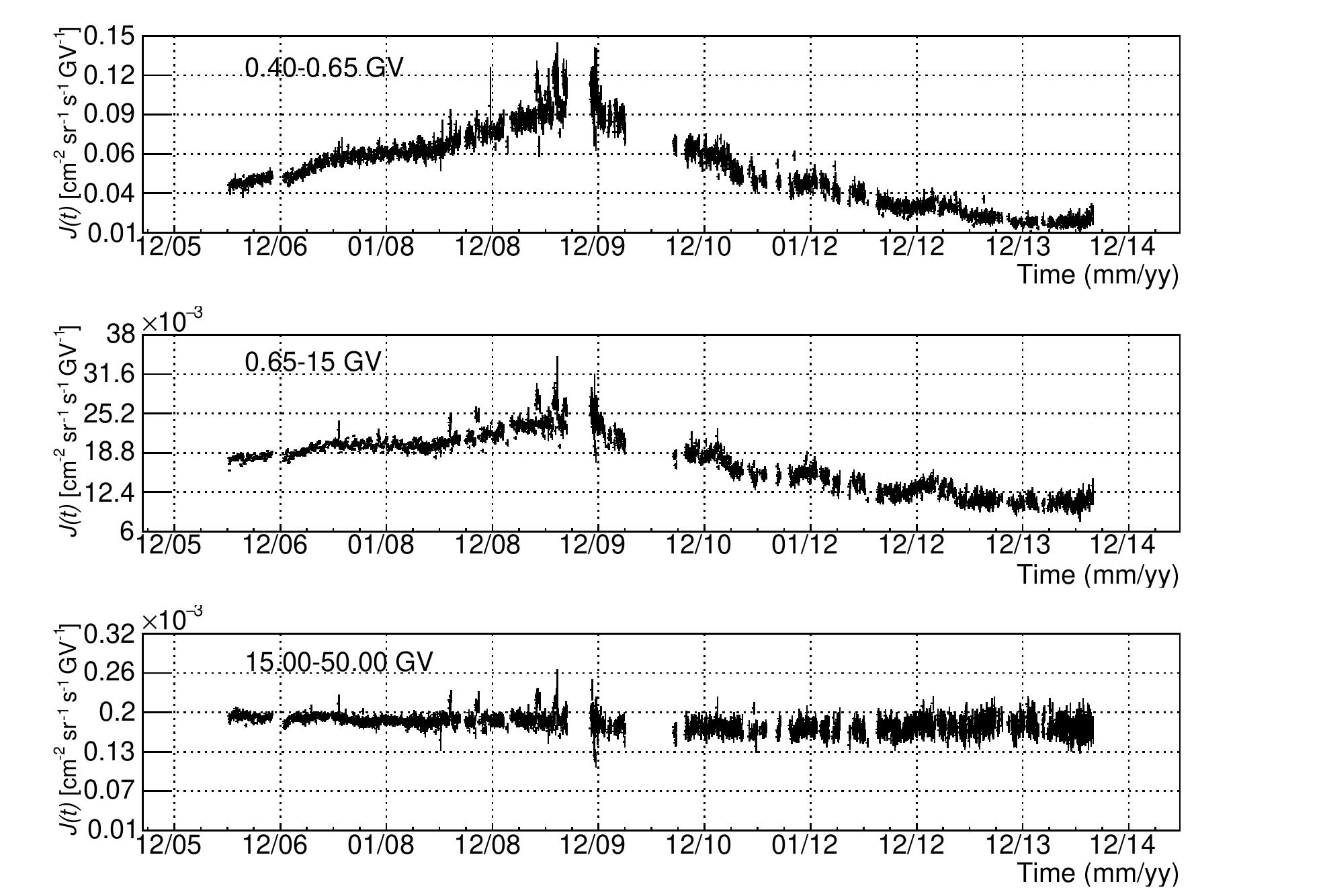}
     \caption{Daily proton intensity $J(t)$ as a function of time for the rigidity intervals described in the text. Different phases of the solar cycles 23$^{rd}$ and 24$^{th}$ are clearly observable. The channel 15-50 GV is almost unaffected by solar modulation.}
     \label{fig:Fig1}
\end{figure}

In Figure \ref{fig:Fig1}, the daily proton intensity time profile $J(t)$ for the aforementioned rigidity intervals is shown. The large gaps in the data (between 2010 and 2011) are due to periods in which the instrument was not fully operational. Different phases of the solar cycle are visible in the shape of the intensity profiles: from 2006 to January 2009 solar cycle 23$^{rd}$ comes to an end and the proton flux slowly increases due to a stable condition of the heliosphere, as already described in \cite{Adriani2013}. After 2010, the activity of the 24$^{th}$ cycle gradually rises and the proton flux decreases accordingly.

In addition to  the expected overall trend, almost disappearing after 15 GV, the low rigidity profiles present some small and regular peaks, mostly during the descending phase of the modulation cycle. 
In order to highlight these peaks, a fit of the proton flux $J(t)$ has been performed to try to disentangle possible high frequencies in the proton data from the well known undecennial modulation.  Two distinct third degree polynomials, in the form $f^3(x)=a+bx+cx^2+dx^3$, one for the data during the ascending phase $J_1(t)$ and  another for the descending one $J_2(t)$, have been used. This approach ensures a statistically good compromise between number of free parameters and precision. 
The fluctuations $\xi_1(t)$ and $\xi_2(t)$ between the experimental data of the two solar phases and the results of the respective fits $f^3(t)$  were evaluated without taking into account the period around the maximu:

\begin{align}
\begin{split}
\xi_1(t)=J_1(t)-f^3(t)\\
\xi_2(t)=J_2(t)-f^3(t).
\label{eq:eq1}
\end{split}
\end{align}

These fluctuations are presented together, as $\xi(t)$, for each of the two most significant rigidity channels in the first two panels of Figure \ref{fig:Fig3}. A quasi-periodic oscillation appears, more evident after December 2010. For comparison, the same technique has been applied to the data of the Apatity Neutron Monitor (\url{http://www.nmdb.eu/nest/search.php/}); the results, reported in the third panel of Figure \ref{fig:Fig3}, show a periodicity that seems to coincide with the observed one by PAMELA.
To try to explore the origin of this quasi-periodicity, a raw periodgram (95\% Kolmogorov-Smirnov confidence level) of the first two rigidity spectra, 0.4-0.65 GV and 0.65-15 GV, has been carried out. The results are presented in the fifth panel of Figure \ref{fig:Fig3}. Peaks appear around 580, 450, 370, 320 and 230 days. The 580-days periodicity is possibly linked to a $\sim$600-days periodicity reported in \cite{Valdes1996}, which could be associated with fluctuations in the southern  coronal hole area and in large active regions. The seasonal $\sim$370-days cosmic ray variation is caused by the Earth's rotation \cite{Forbush1954}. The 230 days peak was also found in \cite{Mohamed_2011} as a 0.7 years periodicity during the A$>$0 solar cycle (1992-2000). The last periodicity, $\sim$450 days, was already described in \cite{Valdes1996} as a possible 1.2 years periodicity but its origin is not reported.


The nature of the $\sim$450 days periodicity could be related to Quasi Biennial Oscillations or QBOs \cite{Vecchio2012,Laurenza2012}. 
QBOs have been detected as a prominent scale of variability in GCRs, but they could just be more an effect of superposition of other periodic/quasi-periodic  processes and not stochastic perturbations.  Higher QBOs have been observed during solar maxima with respect to solar minima \cite{Bazilevskaya2015}; this could explain a higher signal in the descending phase of PAMELA proton data (see Figure \ref{fig:Fig1}) and could be related to different drift effects in different polarities of the HMF. In the past, numerous periodicities between 0.5 and 4 years have been correlated to QBOs \cite{Kato2003,Rybak2001,Kudela2002,Kudela2010,Benevolenskaya1998} and it is very difficult to disentangle their effects form others.

In this work a different hypothesis is proposed for the 450 days periodicity as a study case: a Jovian origin, more exotic but largely proposed in the past. It is well known that the planet Jupiter possesses a very intense magnetosphere due to the combination between its strong magnetic field, about $10^4$ times larger than that of the Earth, and the weakness of the solar wind at 5 AU. 
Evidence that Jupiter could generate high-energy particles has been shown for electrons \cite{Teegarden1974,Simpson1974} and more recent studies revealed that the impulsive and quasi-periodic bursts observed in Jupiter's duskside magnetosphere contain also protons and helium nuclei in the range 0.7-10 MeV/nucleon \cite{Zhang1995}.

Moreover, indication that acceleration mechanisms can operate in magnetospheric environments has been found on Earth, where trapped particles are shown to be accelerated to relativistic energies by local acceleration acting in the heart of the Van Allen radiation belts \cite{Reeves2013}.

\begin{figure}
	\centering
	\includegraphics[width=0.7\textwidth]{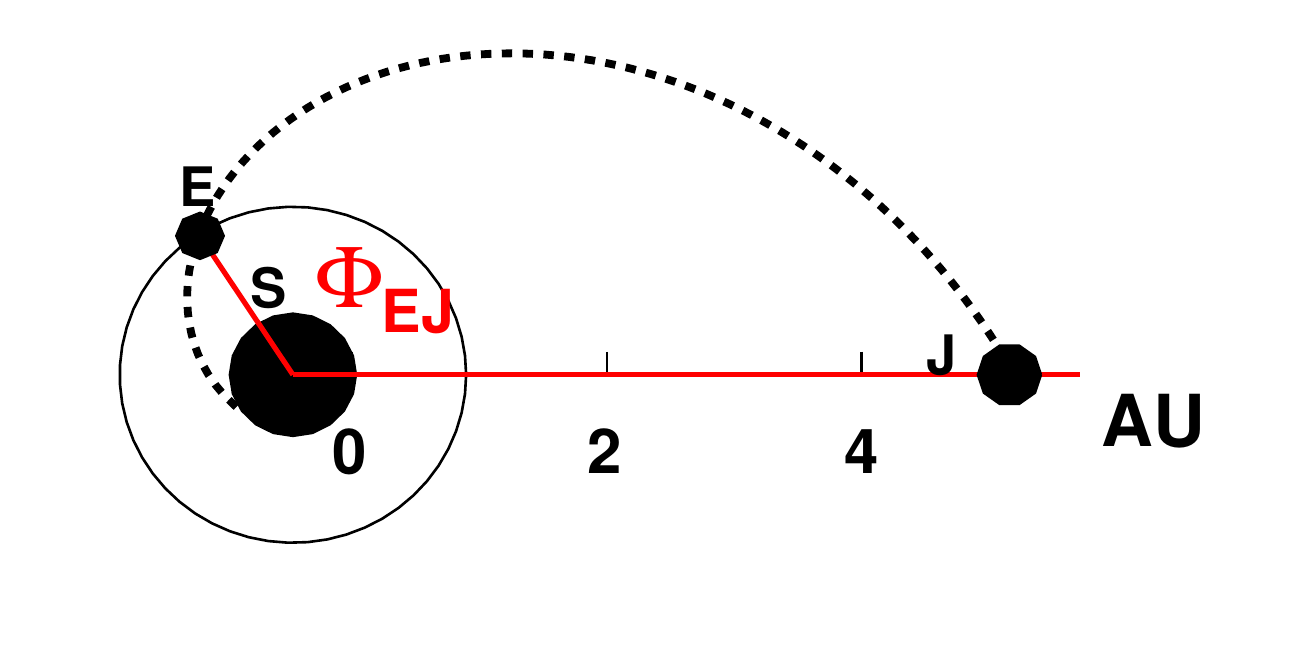}
	\caption{A schematic view of the Earth-Sun-Jupiter system with the IMF line connecting Jupiter with Earth (dashed line). The angle $\Phi_{EJ}$ describes the purely geometrical angle between both planets.}
    \label{fig:FigJup}
\end{figure}

\begin{figure}
  \centering
     \includegraphics[width=\textwidth]{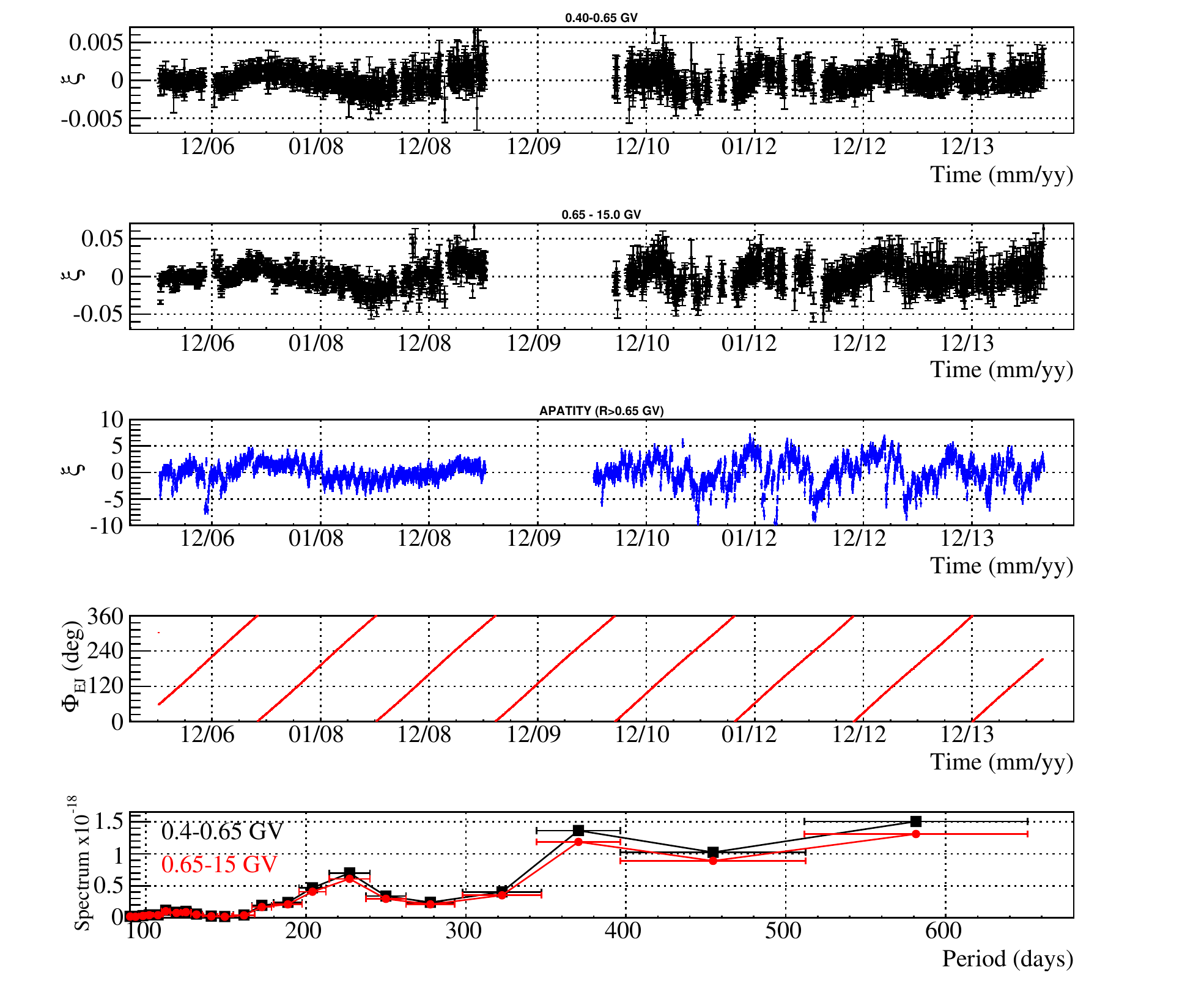}
     \caption{The fluctuations $\xi(t)$ as a function of time are presented together for the two most significant rigidity channels of PAMELA (first two panels). A quasi-periodic oscillation appears before and after the gap, more visible starting from December 2010. In the third panel the same periodicity seems to be also present in the Apatity neutron monitor data. The fourth panel represents the time profile of the angle $\Phi_{EJ}$ between Earth and Jupiter. The last panel shows the results of the  periodgram performed on PAMELA data in the first two rigidity channels (0.4-0.65 GV and 0.65-15 GV).}
     \label{fig:Fig3}
\end{figure}

The hypothesis that some cosmic rays observed at Earth be generated in the Jovian magnetosphere, at least up to energies of the order of few GeV, and then injected in the interplanetary space along the magnetic-field force lines,  has been discussed in the past \cite{Pizzella1973,Pizzella1975,Mitra1983} using the  observations of ground stations. 
With a synodic period of 398.88 days, the Jovian assumption could be somehow related to the $\sim$450 days periodicity found in PAMELA data. 
In fact, having also regard to the uncertainty bar  in the periodgram of Figure \ref{fig:Fig3}, the peak around 450 days could be compatible with a value close to 400 days, but the limited data-taking is not sufficient to clearly resolve them.
 It is worth to note that if some protons arrive guided by the Interplanetary Magnetic Field (IMF) lines connecting Jupiter to Earth, larger fluxes are expected at certain angles.
If an angle $\Phi_{EJ}$ is defined as the Earth longitude in a reference system with center in the Sun and co-rotating with Jupiter, as shown in Figure \ref{fig:FigJup}, it is possible to associate to every daily proton intensity measured by PAMELA the respective value of the geometrical angle $\Phi_{EJ}$, that can be obtained from \url{http://pds-rings.seti.org/tools/ephem2_jup.html}.

The $\Phi_{EJ}$ profile as a function of time is shown in Figure \ref{fig:Fig3}, fourth panel. 

Recent descriptions, based on satellite data, give an IMF with spiral lines of force, but with a precise behavior that depends on the radial gradients of the magnetic field itself. It is found that the favored angle $\Phi_{EJ}$ of the IMF lines connecting Earth and Jupiter is less than 180$^{\circ}$: 140$^{\circ}$ in \cite{Khabarova2012}, 117$^{\circ}$ in \cite{Behannon1978} or 150$^{\circ}$/160$^{\circ}$ in \cite{Mitra1983}. 

Therefore, we followed an approach that takes into account these results.

The de-trended daily proton averages, $\xi(t)$, obtained through Equation \ref{eq:eq1}, have been distributed in two samples: (a) protons associated with angles $\Phi_{EJ}\leq180^{\circ}$ and (b) protons associated with angles $\Phi_{EJ}>180^{\circ}$. 
If some protons arrive from Jupiter, a higher number of protons is expected in sector (a) with respect to sector (b), due to the Archimedean configuration of the Parker spiral originating from the Sun \cite{Parker1976}.

Each of the two samples, (a) and (b),  has further  been  separated in two sections, according to whether the protons were collected during the ascending phase of the solar  modulation cycle or the descending phase:

\begin{itemize}
\item a1 (ascending, $\Phi_{EJ}\leq180^{\circ}$) from 9 July 2006 to 9 July 2009
\item b1 (ascending, $\Phi_{EJ}>180^{\circ}$) from 9 July 2006 to 9 July 2009
\item a2 (descending, $\Phi_{EJ}\leq180^{\circ}$) from 9 July 2010 to 31 August 2014
\item b2 (descending, $\Phi_{EJ}>180^{\circ}$) from 9 July 2010 to 31 August 2014.
\end{itemize}

After that, the average of each sample, $\eta_a$(1,2) and $\eta_b$(1,2), has been evaluated, together with its standard deviation, having verified that the distributions are normal.

Then, a variable $D_j$(j=1,2) is introduced for each couple $\eta_{aj}$(j=1,2) and $\eta_{bj}$(j=1,2) with the respective standard deviation:

\begin{align}
\begin{split}
&D_j~[proton/(cm^2~sr~s~GV)] =\eta_{aj}-\eta_{bj}~~~(j=1,2) \\
&\sigma_{j,total}~[proton/(cm^2~sr~s~GV)]=\sqrt{\sigma_{aj}^2+\sigma_{bj}^2}~~~(j=1,2)
\label{eq:eq2}
\end{split}
\end{align}

The results are given in the Table \ref{tab:risultati} for the three rigidity channels. 

The same procedure  has been followed also  after summing a1 and b1  and  a2 and b2, i.e. taking into account   the entire period of data collection . The results appear in the Table \ref{tab:risultati} labeled as Total. 

For a uniform distribution of the protons along the Earth orbit, a  value of  D$=$0, within errors, is expected. D results always positive with very large Signal/Noise Ratio (SNR = $D/\sigma_{total}$) for each phase and each energy interval. The excess of protons in sector (a) with respect to those in sector (b) is evident especially for the two lowest energy ranges.

\begin{table}
\centering
\begin{tabular}{|cc|cc|c|c|cc|}
\hline
\hline
Rigidity &solar &D&$\sigma_{total}$&excess&SNR&$\chi^2/ndf$&$\chi^2/ndf$\\
interval [GV]&phase&&&\%&&$0-180^o$&$180-360^o$\\
\hline
\hline
0.4-0.65	&total&			 0.000609 & 0.000052 & 4.3 & 11.7&0.97&0.98\\
0.4-0.65	&ascending&			 0.000462 & 0.000067 & 2.6 & 6.9& 0.77 & 0.74\\
0.4-0.65	&descending&			 0.00073 & 0.000076 & 7.4 & 9.6&0.98&1.4\\
\hline
0.65-15	&total	&	    0.005948 & 0.00060 &   2.5&9.9&1.35&0.70\\
0.65-15&ascending	&	 0.002056 & 0.00074 &0.72  &2.8&  0.81&0.93\\
0.65-15&descending&0.00925&0.00091&4.8&10.2&1.1&0.97\\
\hline
15-50	&total	&	     0.00000172 &0.00000041&   0.96&4.2&2.0&1.7\\
15-50&ascending	&	0.00000137 &  0.00000042 &0.74  &3.3&  0.99&1.4\\
15-50&descending&0.00000203&0.00000069&1.2&2.9&1.4& 1.0 \\
\hline
\end{tabular}
\caption{D and $ \sigma_{total}$ in unit of proton/$(cm^2~sr~s~GV)$, the excess of proton flux and SNR=$D/\sigma_{total}$ are represented. The last two columns show the $\chi^2/ndf$ calculated for gaussian fits for the categories (a) and (b) in the text. We remark that all D values are positive and with large SNR.}
\label{tab:risultati}
\end{table}

\begin{figure}
  \centering
    \includegraphics[width=0.8\textwidth]{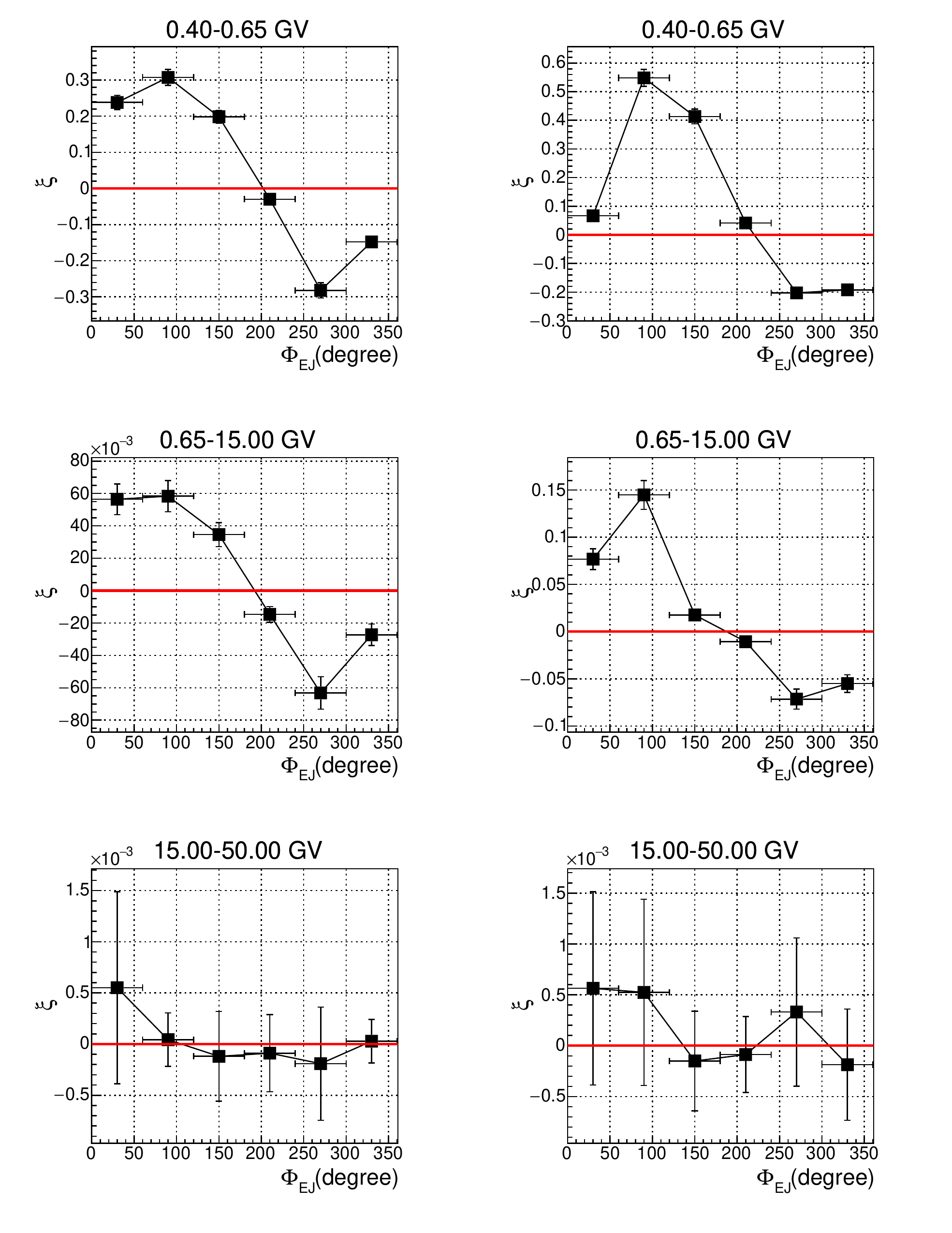}
     \caption{Distribution of the fluctuations $\xi$ as a function of the relative positions of Earth and Jupiter, $\Phi_{EJ}$, for the rigidity channels 0.4-0.65 GV, 0.65-15 GV and 15-50 GV. Left panel refers to the ascending phase, while right panel to the descending one. A maximum around 100$^{\circ}$ is observed in the first two channels.}
     \label{fig:angle}
\end{figure}

If this excess was all due to Jupiter, the contribution to the GCR should be about 4.3\% for the rigidity channel 0.4-0.65 GV, and 2.5\% for the channel 0.65-15 GV.
It is worth to consider that, if part of the protons arrived along the IMF lines from Jupiter, larger fluxes should be expected at certain values of the angle $\Phi_{EJ}$. To explore also this behavior, each value of the fluctuations $\xi$, obtained through Equation \ref{eq:eq1}, has been plotted as a function of the angle $\Phi_{EJ}$ instead of the time (like in Figure \ref{fig:Fig3}). The resulting angular distribution is depicted in Figure \ref{fig:angle} for the three rigidity channels, both for the ascending (left panel) and the descending (right panel) phases of the proton intensities.
It is seen that there is an increase of GCR protons for both the  phases when the angle $\Phi_{EJ}$ lies between 60$^{\circ}$ and 200$^{\circ}$ with a maximum around 100$^{\circ}$. The peak disappears almost entirely in the interval 15-50 GV.

\section{Discussion and conclusions}

It is worth to note that the precise PAMELA measurements, to our knowledge, are the first detected in space-borne experiments, exploring large periodicities in GCRs intensities and providing also rigidity information. 
The results obtained in this work are in good agreement with the previous results reported in \cite{Mitra1983,Pizzella1973} obtained with ground-based detectors. 
The difference with respect to Parker, who found an angle $\Phi_{EJ}\sim$216$^{\circ}$, may be attributed to the IMF configuration considered in the classical calculation, that more recent experiments found to be much more complicated that the one proposed in the past \cite{Khabarova2011,Khabarova2012,Behannon1978}.
It should be also noted that the previous angular results were explained in \cite{Nagashima_1984,Swinson_1974} as a possible effect due to anisotropies linked to the solar cycle.
The results shown in this paper could favor  the idea that the Jupiter magnetosphere might be a source of a small but not negligible fraction of protons measured at 1 AU, accelerating particles  by mechanisms like interaction with the solar wind \cite{Krimigis1981}. 
If this is the right interpretation, we can venture out to conclude that magnetospheres of astrophysical systems (say, Jupiter, Pulsars,...) be possible sources of cosmic rays.



\acknowledgments

We acknowledge partial financial support from The Italian Space Agency (ASI) under the program \``Programma PAMELA - attivit\`a scientifica di analisi dati in fase E\''. We also acknowledge support from Deutsches Zentrum fur Luft- und Raumfahrt (DLR), The Swedish National Space Board, The Swedish Research Council, The Russian Space Agency (Roscosmos) and Russian Science Foundation.

\end{document}